\documentclass[traditabstract, longauth]{aa} 
\usepackage{txfonts}
\usepackage{graphicx}
\usepackage{natbib}
\bibpunct{(}{)}{;}{a}{}{,}
\def\gapprox{{_>\atop{^\sim}}}

\def\cmmd{\rm {cm^{-3}}}
\def\cmmt{\rm {cm^{-2}}}

\def\s-1{\rm {s^{-1}}}

\def\HC3N{HC$_3$N}

\def\kms{\hbox{${\rm km\,s}^{-1}$}}
\def\msun{M$_{\odot}$}
\def\lsun{L$_{\odot}$}

\newcommand{\asec}{\mbox{$''$}}

\usepackage{url}

\begin{document}
 \title{Probing highly obscured, self-absorbed galaxy nuclei with \\ vibrationally excited HCN\thanks{Based on observations
carried out with the IRAM Plateau de Bure and ALMA Interferometers. IRAM is supported by INSU/CNRS (France), MPG (Germany), and IGN (Spain).
ALMA is a partnership of ESO (representing its member states), NSF (USA), and NINS (Japan), together with NRC (Canada) and NSC and ASIAA (Taiwan),
in cooperation with the Republic of Chile. The Joint ALMA Observatory is operated by ESO, AUI/NRAO, and NAOJ.}}


   \author{S. Aalto
          \inst{1}
	\and
          S. Mart\'in\inst{2}
	 \and
	  F. Costagliola\inst{3,1}
          \and
          E. Gonz\'alez-Alfonso\inst{4}
          \and
          S. Muller\inst{1} 
         \and
         K. Sakamoto\inst{5} 
         \and
       G. A. Fuller\inst{6}
       \and
	  S. Garc\'ia-Burillo\inst{7}
        \and	
          P. van der Werf\inst{8}
          \and
	  R. Neri\inst{2}
	\and
         M. Spaans\inst{9}
        \and
         F. Combes\inst{10}
	\and
       S. Viti\inst{11}
	\and
       S. M\"uhle\inst{12}
	\and
	L. Armus\inst{13}
	\and
 	A. Evans\inst{14}
	\and
	E. Sturm\inst{15}
	\and
	J. Cernicharo\inst{16}
 \and
          C. Henkel\inst{17,18}
\and
T.~R.~Greve\inst{19},
             }

 \institute{Department of Earth and Space Sciences, Chalmers University of Technology, Onsala Observatory,
              SE-439 94 Onsala, Sweden\\
              \email{saalto@chalmers.se}
\and Institut de Radio Astronomie Millim\'etrique (IRAM), 300 rue de la Piscine, Domaine Universitaire de Grenoble,
38406 St. Martin d$ ' $H\`eres, France
 \and Instituto de Astrof\'isica de Andaluc\'ia, Glorieta de la Astronom\'ia, s/n, (IAA-CSIC), 18008, Granada, Spain
\and Universidad de Alcal\'a de Henares,Departamento de F\'isica y Matem\'aticas, Campus Universitario, 28871 Alcal\'a de Henares, Madrid, Spain 
\and  Institute of Astronomy and Astrophysics, Academia Sinica, PO Box 23-141, 10617 Taipei, Taiwan 
\and Jodrell Bank Centre for Astrophysics, School of Physics \& Astronomy, University of Manchester, Oxford Road, Manchester M13 9PL, UK
       \and Observatorio Astron\'omico Nacional (OAN)-Observatorio de Madrid, Alfonso XII 3, 28014-Madrid, Spain
       \and Leiden Observatory, Leiden University, 2300 RA, Leiden, The Netherlands
\and Kapteyn Astronomical Institute, University of Groningen, PO Box 800, 9700 AV Groningen, The Netherlands 
\and Observatoire de Paris, LERMA (CNRS:UMR8112), 61 Av. de l'Observatoire, 75014 Paris, France 
\and Department of Physics and Astronomy, UCL, Gower St., London, WC1E 6BT, UK 
\and Argelander-Institut f\"ur Astronomie, Auf dem H\"ugel 71, 53121 Bonn, Germany
\and Infrared Processing and Analysis Center, California Institute of Technology, 1200 East California Boulevard, Pasadena, CA 91125, USA 
\and University of Virginia, Charlottesville, VA 22904, USA, NRAO, 520 Edgemont Road, Charlottesville, VA 22903, USA
\and Max-Planck-Institut f\"ur extraterrestrische Physik, Postfach 1312, 85741 Garching, Germany 
\and Group of Molecular Astrophysics, ICMM, CSIC, C/Sor Juana Ines de La Cruz N3, E-28049 Madrid, Spain
\and Max-Planck-Institut f{\"u}r Radioastronomie, Auf dem H{\"u}gel 69, 53121 Bonn, Germany
 \and Astronomy Department, King Abdulaziz University, P.O. Box 80203
    Jeddah 21589, Saudi Arabia
\and Department of Physics and Astronomy, University College London, Gower Street, London WC1E 6BT, UK }

   \date{Received xx; accepted xx}

  \abstract{We present high resolution (0.\asec4) IRAM PdBI and ALMA mm and submm observations of the (ultra) luminous infrared galaxies ((U)LIRGs) IRAS17208-0014,  Arp220, IC860 and Zw049.057 that reveal intense line emission from vibrationally excited ($\nu_2$=1) $J$=3--2 and 4--3 HCN. The  emission is emerging from buried, compact ($r<$17-70 pc) nuclei that have very high implied mid-infrared surface brightness $>$$5\times 10^{13}$ \lsun\ kpc$^{-2}$. These nuclei are likely powered by accreting supermassive black holes (SMBHs) and/or hot ($>$200 K) extreme starbursts.  Vibrational, $\nu_2$=1, lines of HCN are excited by intense 14 $\mu$m mid-infrared emission and are excellent probes of the dynamics, masses, and physical conditions of (U)LIRG nuclei when H$_2$ column densities exceed $10^{24}$ $\cmmt$.  It is clear that these lines open up a new interesting avenue to gain access to the most obscured AGNs and starbursts. 
Vibrationally excited HCN acts as a proxy for the absorbed mid-infrared emission from the embedded nuclei, which allows for reconstruction of the intrinsic, hotter dust SED.
In contrast, we show strong evidence that the ground vibrational state ($\nu$=0), $J$=3--2 and 4--3 rotational lines of HCN and HCO$^+$ fail to probe the highly enshrouded, compact nuclear regions owing to strong self- and continuum absorption.  The HCN and HCO$^+$ line profiles are double-peaked because of the absorption and show evidence of non-circular motions - possibly in the form of in- or outflows.  
Detections of vibrationally excited HCN in external galaxies are so far limited to ULIRGs and early-type spiral LIRGs, and we discuss possible causes for this.  We tentatively suggest that the peak of vibrationally excited HCN emission is connected to a rapid stage of nuclear growth, before the phase of strong feedback. 
  }


 \maketitle


\section{Introduction}

Luminous  ($L_{\rm IR}$=$10^{10}-10^{11}$ \lsun) and ultraluminous ($L_{\rm IR}\gapprox 10^{12}$ \lsun) infrared galaxies ((U)LIRGS) are powered by either bursts of star formation or AGNs (active galactic nuclei - accreting supermassive black holes (SMBHs)) and are fundamental to galaxy mass assembly over cosmic time \citep[e.g.][]{elbaz03,sanders96}.  Some
(U)LIRGS have deeply embedded nuclei that harbour a very active evolutionary stage of AGNs and starbursts, often with signatures of outflowing and infalling
gas \citep[e.g.][]{banerji12,fabian99,gonzalez12,silk98, spoon13}. These nuclei are key to understanding nuclear growth and feedback mechanisms
and studying them is essential for a complete AGN and starburst census, for constraining orientation-based unification models and fully understanding how galaxies grow and evolve \citep[e.g.][]{brightman12, merloni14}.

However, probing inside the optically thick layers of the most enshrouded galaxy nuclei  is an observational challenge.
X-ray and mid-infrared (mid-IR) surveys are designed to find embedded nuclear activity, but when $A_{\rm v} >1000$ magnitudes ($N$(H$_2$)$>10^{24}$ $\cmmt$), the emission becomes extremely attenuated even at these wavelengths \citep[e.g.][]{treister10,lusso13, spoon02, roche15}. Continuum imaging at the more transparent radio- \citep[e.g.][]{parra10} and mm/sub-mm \citep[e.g.][]{sakamoto08} wavelengths can be used to probe luminosity density and to identify power sources, although interpretations are sometimes complicated by surrounding star formation, free-free absorption, and other optical depth effects. Rotational lines (in the vibrational ground state $\nu$=0) of polar molecules, such as HCN and HCO$^+$, are common tracers of dense ($n>10^4$ $\cmmd$) gas and are often used to probe the nuclear mass, dynamics, and astrochemistry \citep[e.g.][]{gao04,meijerink05} of obscured galaxy nuclei. However, large optical depths and the low-level energies of the transitions (see Fig.~\ref{f:energy}) make them less well suited to studying deeply embedded, hot, and compact starbursts or obscured AGNs.


\medskip
\noindent

In contrast, rotational transitions of vibrationally excited ($\nu_2$=1) HCN occur between energy levels exceeding 1000 K above ground (Fig.\ref{f:energy}) \citep{ziurys86}. 
These transitions are usually radiatively excited by the intense mid-IR emission provided by the presence of hot dust and high column densities (see Sect.~\ref{s:intro2} and Appendix~\ref{s:B}).
Therefore, the dust and gas obscuration impeding the use of other tracers is
a {\it prerequisite} for the excitation of vibrational HCN.  Since the rotational  transitions of HCN occur in the mm and sub-mm regimes the line emission can better penetrate the optically thick layers of dust, so it functions as an unattenuated tracer of deeply buried high mid-infrared surface brightness nuclei. 

\smallskip
\noindent
Extragalactic vibrationally excited HCN has been detected in the LIRG NGC4418 \citep{sakamoto10} and the ULIRGs Arp220 \citep{salter08},  IRAS20551-4250 \citep{imanishi13},
and Mrk231 \citep{aalto15}. In NGC4418 and Arp220, vibrational lines of HC$_3$N have also been used to study the properties of the buried nucleus 
\citep{costagliola10, sakamoto10, martin11, costagliola13, sakamoto13, costagliola15}. Vibrational temperatures ranging from 200 to 400 K have been estimated for HCN and HC$_3$N in
NGC4418, Arp220 and Mrk231. Vibrationally excited HCN has been detected in the circum-nuclear disk at the centre of the Milky Way by \citet{mills13}, but line luminosities are significantly
lower (relative to those of $\nu$=0 HCN) compared to those in the galaxies studied here.

\medskip
\noindent
In this paper we present IRAM Plateau de Bure A-array observations of $J=3\to 2$, $\nu$=0 HCN, HCO$^+$ and  $\nu_2$=1 HCN emission in the LIRGs IC860 and Zw049.057.  
We also present Atacama Large Millimeter Array (ALMA) observations of the corresponding $J=4\to 3$ lines in the ULIRGs Arp220 and IRAS17208-0014. 
These new observations double the number of galaxies with detected vibrationally excited HCN emission and reveal in unprecedented detail emission emerging from highly compact structures behind a 
 temperature gradient. We can show, for the first time, that the ground vibrational state HCN and HCO$^+$ line emission becomes self- and continuum absorbed and fails to trace the inner, nuclear regions of these galaxies. 

\bigskip
\noindent

\section{Vibrationally excited HCN}
\label{s:intro2}

\begin{figure}
\includegraphics[scale=0.5]{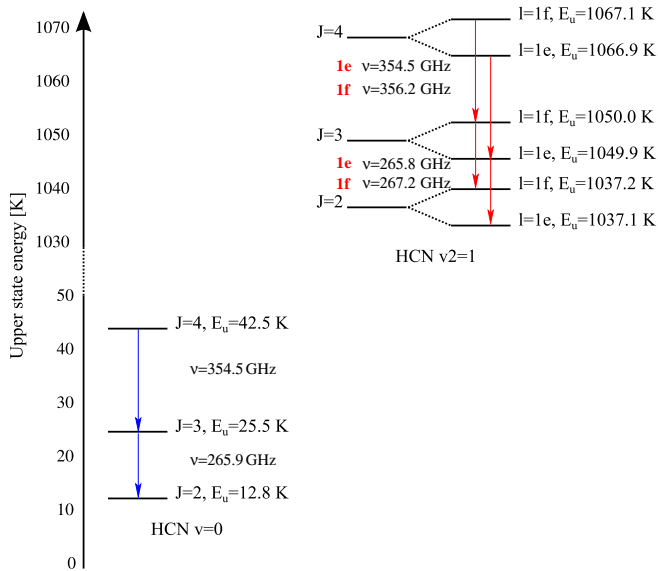}
\caption{\label{f:energy} HCN energy diagram with the HCN $\nu$=0 and $\nu_2$=1 $J$=4,3,2  levels marked. The split of the  $\nu_2$=1 rotational levels into $e$ and $f$ is
also shown. The energy levels are not to scale which is illustrated by the dashed y-axis. The transitions are indicated with arrows. The values for the energy levels are from the Cologne
Database for Molecule Spectroscopy (CDMS) (https://www.astro.uni-koeln.de/cdms)
}
\end{figure}

The HCN molecule has degenerate bending modes in the IR and may absorb IR-photons to its first vibrational state (Fig.~\ref{f:energy}). The first excited bending state of
HCN is doubly degenerate, so that each rotational level $\nu_2$=1 $J$ is split into a closely spaced pair of levels ($f$ and $e$). 
The $J$=3--2 and 4--3 $\nu_2$=1
lines are rotational transitions within the vibrationally excited state that occur near the frequency of the corresponding
$\nu$=0 lines. For example, the $\nu_2$=1$e$ 3--2  line is shifted by only 20 MHz from the $\nu$=0 HCN 3--2  line and therefore appears close to the centre of the
HCN line for extragalactic sources while the $\nu_2$=1$f$ line has rest frequency $\nu$=267.1993 GHz, shifted by 358 MHz (+400 \kms) from the HCO$^+$ 3--2 line.  
The $\nu_2$=1$e$ and 1$f$ line emission always have similar intensity (except for cases of population inversion).

\smallskip
\noindent
We will hereafter refer to the HCN $\nu$=0 lines as ”HCN” and the vibrationally excited HCN $\nu_2=1f$ lines as ”HCN-VIB”.

\subsection{Mid-IR excitation of the HCN-VIB lines}

The energy above ground of the $\nu_2$=1 state starts at $T_{E/k}$=1024 K and the critical density of the HCN-VIB lines is large enough ($>10^{10}$ $\cmmd$) that
collisional excitation is unlikely \citep{ziurys86,mills13}. Instead, the vibrational ladder
can be excited through the absorption of mid-IR 14$\mu$m continuum emission. There is no sharp threshold, but at mid-IR background temperatures of
$T_{\rm B}$(14$\mu$m)$\approx$100 K the  pumping is efficient enough to start to excite the line \citep{carroll81,aalto07}. 

For a thorough investigation of the vibrationally excited HCN a multi-line, non-LTE analysis is required. For example, if the buried source is very hot, compact and luminous it may
be necessary to also consider higher vibrational modes  ($\nu_1$ at 3.02$\mu$m and $\nu_3$ at 4.77$\mu$m) for the excitation of  HCN-VIB \citep{cernicharo11}. In a
forthcoming paper we will present non-LTE models of HCN-VIB emission excited by buried IR sources (Gonz\'alez-Alfonso et al. (in prep.)). 

However, for the purpose of this discussion we will
assume that HCN-VIB is excited primarly by 14$\mu$m continuum with  a brightness temperature $T_{\rm B}\gapprox$100 K.


\section{Observations}
\label{s:obs}

\subsection{Sample galaxies}

The studied galaxies (see Tab.~\ref{t:gal}) are part of a sample where we find evidence of compact nuclear activity with faint 158 $\mu$m [C II] line emission relative
to FIR continuum \citep{malhotra97,diaz13} and/or that show prominent 14$\mu$m HCN absorption \citep{lahuis07}. The [C II] deficit correlates with degree of
compactness of the IR emitting source and the $L_{\rm FIR}$/$M_{{\rm H}_2}$ \citep{diaz13}.

The studied galaxies also have unusually bright HC$_3$N line emission which we previously have suggested is due to a compact buried hot nucleus
\citep{aalto07, costagliola10, costagliola11}.  Note that we only include the western nucleus of the ULIRG merger Arp220 in this study (hereafter Arp220W)
which is the brighter of the two nuclei.  The eastern nucleus is discussed in Mart\'in et al. (in prep.)

\subsection{IRAM Plateau de Bure}

\begin{table*}[tbh]
\caption{\label{t:gal} Sample galaxies$^{\dag}$.}
\begin{tabular}{lllcccccc}
\hline
\hline\\
\\
Name & R.A. & Dec. & Log $L_{\rm FIR}$ & $D$ & Type & Telescope & Beam & Sensitivity$^{\star}$ \\
& (J2000) & (J2000) & [$L_{\odot}$] & [Mpc] & &&($\asec$) & (mJy / channel) \\
\hline
\\
IC860 &                   13:15:03.05 &   +24:37:08.0  & 11.17 & 59 & Sab & IRAM PdBI & $0.\asec 50 \times 0.\asec 23$ & 1.2 mJy \\
Zw049.057 &           15:13:13.10 &   +07:13:32.0 & 11.27 & 59 & Sab & IRAM PdBI & $0.\asec 55 \times 0.\asec 24$ & 1.6 mJy  \\
Arp220W &               15:34:57.27 &    +23:30:10.5 & 12.03 & 80 & merger & ALMA & $0.\asec 44 \times 0.\asec 38$ & 1.9 mJy  \\
IRAS17208$-$0014 & 17:23:21.94 & $-$00:17:01.0 & 12.39 & 176 & merger & ALMA & $0.\asec 49 \times 0.\asec 34$ & 1.6 mJy \\
\hline
\end{tabular}
\\
\newline
\\
\begin{minipage}[h]{0.9\hsize}
$^{\dag}$$L_{\rm IR}$ and $D$ from \citet{sanders03} (where $D$ is the angular distance ($D_A$)) but for Arp220W we attempt to estimate the luminosity emerging from the western nucleus
only from the continuum study of \citet{sakamoto08}. 
$^{\star}$ For channel widths of 33 \kms (IC860 and Zw049.057) and 20 \kms (Arp220W and IRAS17208-0014).
\end{minipage}

\end{table*}

The 1~mm observations were carried out with the 6-element array in March 2014 in A-array.
The receivers were tuned to a frequency of 263.0~GHz, to include both the redshifted HCN and HCO$^+$ $J=3-2$ lines
($\nu$=265.886~GHz and 267.558~GHz rest frequency, respectively). We used the WideX correlator providing a broad frequency
range of 3.6~GHz and 2~MHz spectral resolution.  After calibration within the GILDAS reduction package, the visibility set was converted
into FITS format, and imported into the AIPS package for further imaging.  We adopted natural weighting during the CLEAN deconvolution.
Table~\ref{t:gal} gives the beam sizes and sensitivities of the final data. The details of the observations are:

\paragraph{IC860:} 
The bandpass of the individual antennas was derived from the bright quasar 3C84 and the absolute flux calibration was set on MWC349.
The quasars J1310+323 ($\sim 0.7$~Jy at 1~mm) and J1328+307 ($\sim 0.2$ Jy) were observed regularly for complex
gain calibration every 25 minutes.

\paragraph{Zw049.057:} 
The bandpass and flux calibrations were derived from 3C279 and MWC349, respectively. The gain calibration was derived from observations
of J1502+106 ($\sim 0.4$ Jy) and J1456+044 ($\sim 0.2$ Jy) with a similar scheme as for IC860.

\subsection{ALMA}
The final sensitivity and beam sizes of the observations are given in Table~\ref{t:gal} and the details are discussed below.

\paragraph{IRAS17208-0014:} The 0.8~mm observations were carried out with ALMA (with 34 antennas in the array) in May 2014.
The correlator was set up to cover two contiguous bands of 1.875~GHz in each of the lower (centered at $\sim$341.9~GHz) and upper
(centered at $\sim$352.1~GHz) sideband, with 960 spectral channels (1.95~MHz wide each) per band. The upper side band includes the
HCN $J=4-3$ $v=0$ and $v=1f$ lines.
The bandpass of the individual antennas was derived from the bright quasar J1733-1304. The quasar J1751+096 ($\sim 1.7$~Jy)
was observed regularly for complex gain calibration. Its flux was extracted from the ALMA flux-calibrator database for absolute flux calibration.
After calibration within the CASA reduction package, the visibility set was imported into the AIPS package for further imaging.

\smallskip
\noindent
\paragraph{Arp220:} The 0.8~mm observations were carried out with ALMA in April and May 2014 (with 34 and 30  usable antennas, respectively). 
Spectral features presented here were observed in the USB with
$2\times1.875$~GHz spectral windows centered in 347.8 and 349.6~GHz. The bandpass of the individual antennas was derived from the quasar
J1550+0527. The flux calibration was set on Titan,  where intrinsic atmospheric lines were masked. The quasar J1516+1932 ($\sim 0.4$ Jy at 0.9 mm)
were observed regularly for complex gain calibration every 5 minutes.
Calibration, deconvolution and imaging were carried out in the CASA reduction package.



\begin{figure*}
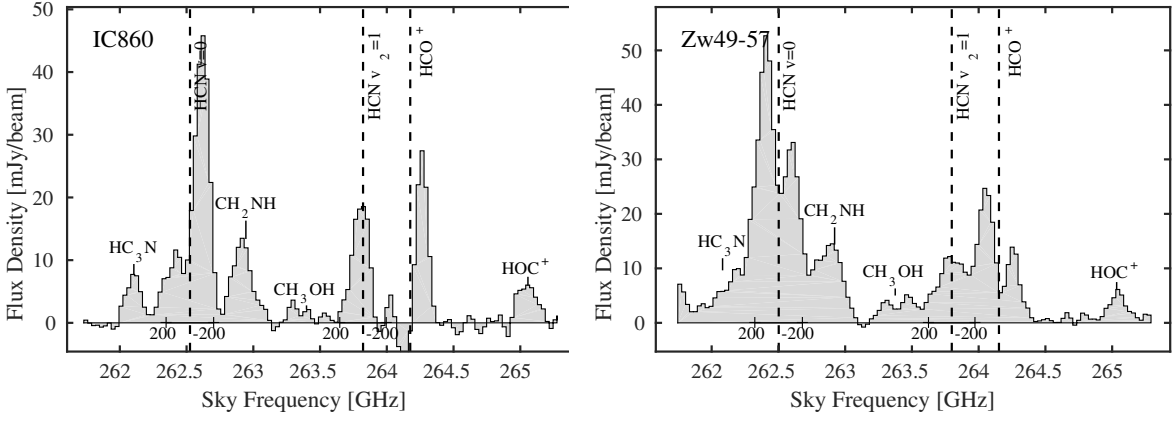

\includegraphics[scale=0.55]{plot3.eps}
\includegraphics[scale=0.55]{plot2.eps}
\caption{\label{f:spec1} IRAM Plateau de Bure spectra  (continuum subtracted) of  IC860 (left panel) and Zw049.057 (right panel). The systemic velocities of HCN, HCO$^+$ and HCN-VIB $J$=3--2 are
marked with dashed vertical lines. The self- and continuum absorption leads to double-peaked HCN and HCO$^+$ $\nu$=0 spectra  while the HCN-VIB 
lines are singled peaked close to systemic velocity (see Sect.~\ref{s:vib}). The spectra also show prominent, single peaked lines of HOC$^+$, CH$_2$NH, CH$_3$OH and HC$_3$N. Fluxes are in mJy beam$^{-1}$ 
}
\end{figure*}

\begin{figure*}
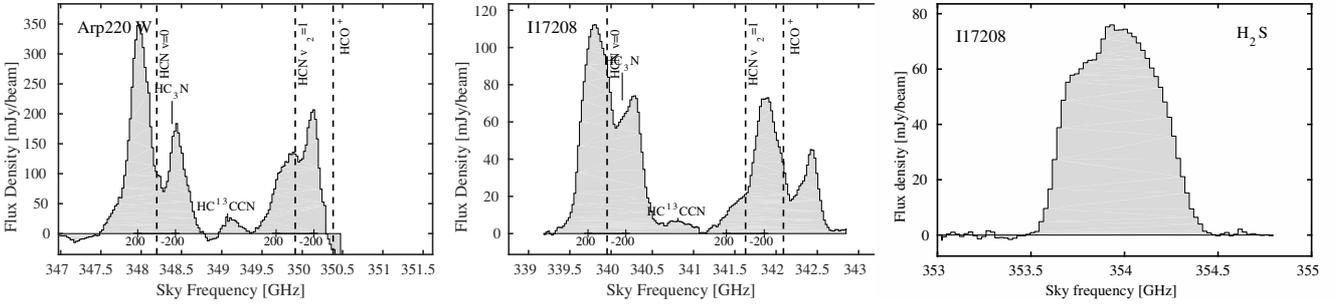

\includegraphics[scale=0.4]{plot1.eps}
\includegraphics[scale=0.4]{plot4.eps}
\includegraphics[scale=0.4]{i17208_h2s.eps}
\caption{\label{f:spec2} ALMA spectra   (continuum subtracted) of Arp220W (left panel) and IRAS17208-0014 (centre and right panels). The systemic velocities of
HCN, HCO$^+$ and HCN-VIB 
$J$=4--3 and the H$_2$S lines are marked with dashed vertical lines. The self- and continuum absorption leads to double-peaked HCN and HCO$^+$ $\nu$=0 spectra 
while the HCN-VIB  and H$_2$S lines are singled peaked close to systemic velocity (see Sect.~\ref{s:vib}). Half of the HCN-VIB is blended with the HCO$^+$ line. Note that
part of the HCO$^+$ line is outside the band for Arp220W.  We also detect lines of HC$_3$N and HC$^{13}$CCN.  Fluxes are in mJy beam$^{-1}$ 
}
\end{figure*}


\section{Results}


Line fluxes, luminosities and source size are presented in  Tab.~\ref{t:flux}. Nuclear (central beam) spectra in Fig.~\ref{f:spec1} and Fig.~\ref{f:spec2}.

\subsection{HCN and HCO$^+$}
\label{s:lines}
HCN and HCO$^+$ source diameters (FWHM) range between 100 and 400 pc (Tab.~\ref{t:flux}) and the line profiles have a striking
double-peaked shape with a minimum near systemic velocity (see Figs.~\ref{f:spec1} and ~\ref{f:spec2}).

\subsubsection{Line shapes} 

In Sect.~\ref{s:self2} we present strong evidence that the double-peaked HCN and HCO$^+$ spectra are caused by deep self- and continuum absorption at systemic velocity.  
Therefore, FWHM widths cannot be fitted for these lines.  In IC860, the shapes of the HCN and HCO$^+$ spectra appear as {\it reversed  P-Cygni profiles} - strongly asymmetric with the red part of the line profile showing stronger absorption. This is particularly striking for HCO$^+$ where the continuum is also absorbed on the red side. The deepest absorption feature occurs at a projected velocity of 70 \kms.    
The emission peak observed between the HCN-VIB line and the HCO$^+$ absorption feature is part of the unabsorbed high velocity HCO$^+$ (similar to what is seen in HCN where the absorption is less pronounced). 

\noindent
In the other galaxies the absorption profiles are instead somewhat blue-shifted by -50 to -100 \kms\ for Zw049.057 and -100 to -300 \kms\ for Arp220W and IRAS17208-0014.

\subsection{Vibrationally excited HCN}
\label{s:vib}

We find luminous HCN-VIB $J$=3--2 or 4--3 1$f$ line emission in all four sample galaxies. The emission originates from the very nucleus of the galaxies with upper limit diameters (FWHM) ranging from 37 to 136 pc (Tab.~\ref{t:flux}).  For the ULIRG merger IRAS17208-0014 the HCN-VIB emission appears to be associated primarily with the western of the two nuclei found by  \citet{medling14}. For all four galaxies, the HCN-VIB emission is significantly more compact (by factors of at least 2-3) than that of HCN and HCO$^+$ and {\it for IC860 the integrated strength of the line exceeds that of HCO$^+$ in the nucleus}.

\begin{figure}
\resizebox{8cm}{!}{\includegraphics[angle=0]{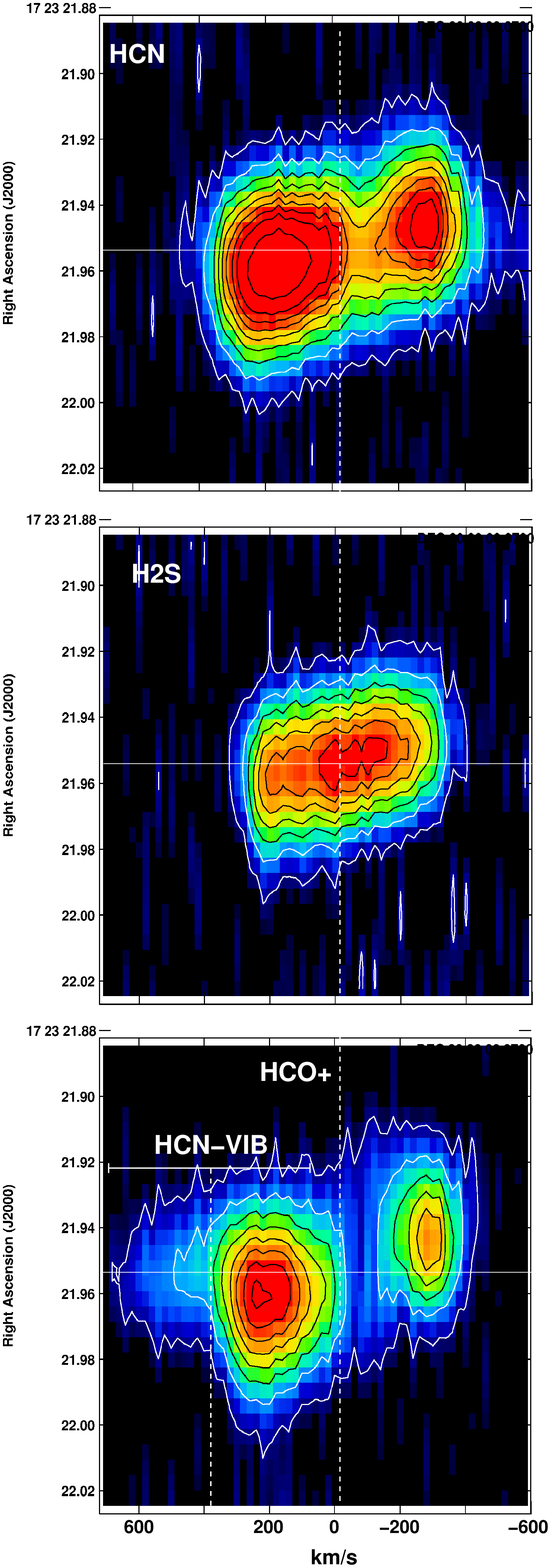}}
\caption{\label{f:I17208pv} Position velocity (pV) diagrams  (continuum subtracted) for IRAS17208-0014 where the y-axis shows the position along the major axis (position angle 90$^{\circ}$) and the
x-axis shows the radial velocity   relative to the $cz$ of 17208-0014 \kms. Upper panel: HCN 4--3. Centre panel: H$_2$S Lower panel:  HCO$^+$ 4--3 blended with HCN-VIB. Colour ranges from 
2 to 70~mJy, contour levels are 5$\times$(1,3,5,...17) mJy beam$^{-1}$.   The vertical dashed line marks $z$=0.04271 the suggested redshift by \citet{garcia15}, while zero velocity marks
the redshift $z$=0.04281.  The horizontal white line indicates the nuclear position of the HCN-VIB line.
}
\end{figure}

The HCN-VIB line widths range from 130 to 450 \kms\ and the emission peaks close to systemic velocity. This is most clearly seen in the two LIRGs where the line blending with HCO$^+$ is less significant.
The centre (systemic) velocities are marked with a dashed line in Figs.~\ref{f:spec1} and~\ref{f:spec2} and for IC~860 it is estimated to $cz$=3880$\pm$20 \kms\ from the position velocity (pV) diagram. The centre velocity for Zw049.057 is very similar  $cz$=3900$\pm$20 \kms\ where the errors also stem from the pV diagram (see Sect.~\ref{s:pv}). For
Arp220W the line marks $cz$=5434 \kms which is 17 \kms\ below the systemic velocity estimated for Arp220W by \citet{sakamoto08}.  It is suggested that the systemic velocity of IRAS17208-0014 is $cz$=12834 \kms\ \citep{downes93}(also marked with a dashed line in Fig.~\ref{f:spec2}) while \citet{garcia15} suggest that the systemic velocity is
blue-shifted by 30 \kms from this value. We have indicated this lower value of the centre velocity in the pV diagram for IRAS17208-0014 (Fig.~\ref{f:I17208pv}).


\begin{figure*}
\resizebox{8cm}{!}{\includegraphics[angle=0]{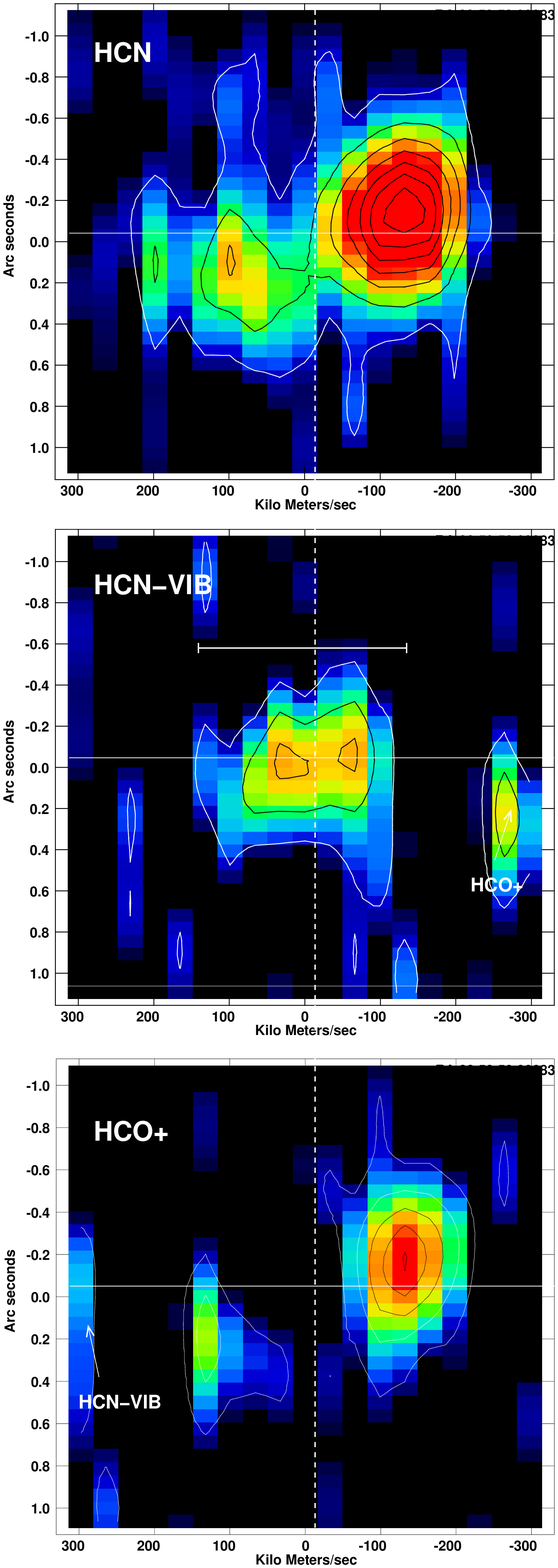}}
\resizebox{8cm}{!}{\includegraphics[angle=0]{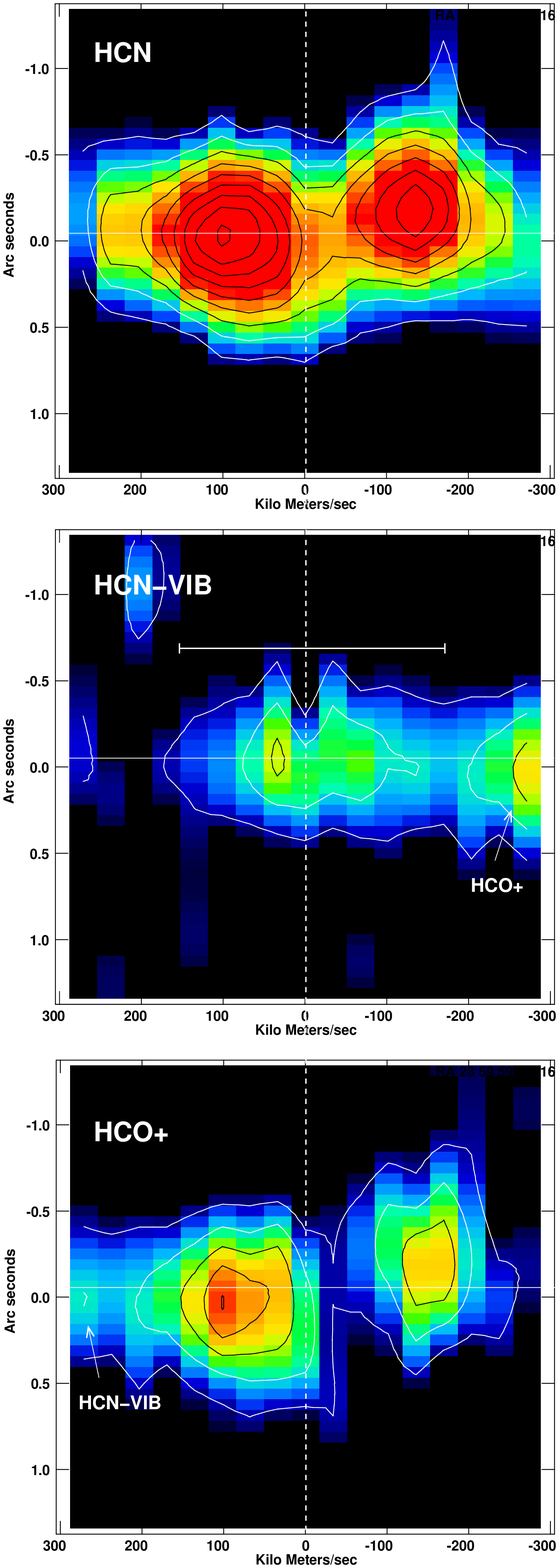}}
\caption{\label{f:pv} Position velocity (pV) diagrams  (continuum subtracted) for IC860 (left) and Zw049.057 (right) where the x-axis shows the radial velocity (relative to $cz$=3880 \kms and 3900 \kms respectively) and the y-axis shows the position along the major axis (for both galaxies the position angle is -20$^{\circ}$). The upper panel shows the HCN, the centre the
HCN-VIB and the lower panel the HCO$^+$ pV diagrams. The contours are  3.5$\times$(1,3,5,7,9,11,13) mJy beam$^{-1}$(IC860) and 
3.5$\times$(1,3,5,...17) mJy beam$^{-1}$ (for Zw049.057). The colour ranges from 2 to 30 mJy beam$^{-1}$.  The white vertical dashed line indicate the systemic velocity. The horizontal
white line indicates the nuclear position of the HCN-VIB line. }
\end{figure*}

\subsection{Other species}

For IC860 and Zw049.057 (see Fig.~\ref{f:spec1}) we detect the HC$_3$N $J$=29--28 $\nu_7$=2 line
($T_{{E_{\rm l}}/k}$=819 K)  on the redshifted and the ($4_{1,3} - 3_{1,2}$) transition of CH$_2$NH at 266 GHz (with energy level $T_{{E_{\rm l}}/k}$=27 K) 
on the blue side of the HCN 3--2 line and several blended CH$_3$OH  lines between CH$_2$NH and HCN-VIB. It is interesting to note that \citet{salter08} find
bright lines of CH$_2$NH in their cm-wave spectral scan of Arp220. We also detect very bright HOC$^+$ 3--2 line
emission on the blue side of the HCO$^+$ 3--2 line.

For Arp220W and IRAS17208-0014 (see Fig.~\ref{f:spec2}) bright $J$=39--38 HC$_3$N line emission is blended with the HCN 4--3 line and
H$^{13}$CCCN is found in between the HCN and HCN-VIB line. For IRAS17208-0014 we have also detected luminous line emission from the $3_{2,1} -3_{1,2}$
and $4_{3,1}-4_{2,2}$ transitions of ortho-H$_2$S  and para-H$_2$S at 369 GHz. The two lines are shifted only 26 MHz from each other (energy levels
$T_{{E_{\rm l}}/k}$=137 K and 245 K respectively) so they are effectively blended into one line.


\subsection{Position velocity diagrams}
\label{s:pv}


For  IC860 and Zw049.057 the HCN-VIB emission is spatially unresolved and peaks at velocities close to systemic (Fig.~\ref{f:pv}). In comparison, the brightest HCN emission occurs outside the HCN-VIB at higher absolute velocities, with a velocity difference between the two peaks of 230-260 \kms.  Even though much less pronounced than for the HCN and HCO$^+$ lines, the HCN-VIB line also seems to be slightly double-peaked. The velocity shift between the peaks is relatively small: $\approx$80 \kms.

For the ULIRGs Arp220W and IRAS17208-0014 the HCN-VIB line is broad enough for it to be partially blended with the HCO$^+$ line (see example for IRAS17208-0014 in Fig.~\ref{f:I17208pv}). Observations at higher spatial resolution are necessary to completely disentangle the HCN-VIB emission.  The HCN-VIB pV diagrams for both Arp220W (to be presented in Mart\'in et al. (in prep.)) and IRAS17208-0014  (Fig.~\ref{f:I17208pv}) therefore only show half of the line but in both cases a single peak at systemic velocity is suggested. For IRAS17208-0014 the 369 GHz H$_2$S line has a profile that does not show the pronounced double structure of HCN and HCO$^+$. The pV diagram (Fig.~\ref{f:I17208pv}) reveals that the H$_2$S line emission peaks where HCN and HCO$^+$ have a minimum. Note also the blue-shifted high-velocity extension to the HCN
line which may be the HCN counterpart to the CO 2--1 high velocity outflow detected by \citet{garcia15}. It may be similar to the HCN 3--2 outflow already found in the ULIRG Mrk–231
by \citet{aalto15}.


\subsection{Continuum}
\label{s:cont}
Continuum fluxes and FWHM source sizes at 260 or 350 GHz can be found in Tab.~\ref{t:cont}.
The FWHM size of the continuum emission region is larger than the HCN-VIB emission for two sources - Zw049.057 and IRAS17208-0014. In IC860 the
continuum source is unresolved as is the HCN-VIB emission. \citet{sakamoto08} and \citet{wilson14} have shown the existence of an optically
thick and compact submm continuum source in Arp220W consistent with our result.  In both IRAS17208-0014 and IC860 the lower limit to the
brightness temperature of the 260 and 350 GHz continuum sources is $T_{\rm B} >$40 K suggesting that these sources may also be optically thick.



\begin{table*}
\caption{\label{t:flux} Line flux densities.}
\begin{tabular}{lcccccc}
\hline
\hline\\
Line & Peak$^{\dag}$  & Integrated$^{\dag}$ & $L$ & $\theta$$^{\dag}$  & PA$^{\dag}$ & $\Delta$V\\
& [Jy\,\kms beam$^{-1}$] & [Jy\,\kms] & [\lsun $\times 10^3$] & [pc] & [$^{\circ}$] & [\kms]\\
\hline \\ 
{\bf IC860}\\
{\it HCN 3--2} & 8.8 $\pm$ 0.3 &  19.4 $\pm$ 0.7 & 18.5 & $128 \times 80$  & 20 & $\dots$   \\
{\it HCN-VIB 3--2} & 3.8 $\pm$ 0.4 &  4.0 $\pm$ 0.4 & 3.8 & $<$37  & $\dots$  & 130\\
{\it HCO$^+$ 3--2} & 3.3 $\pm$ 0.2 &  9.0 $\pm$ 0.4 & 8.6 & $133 \times 94$ & 20 & $\dots$   \\ 
{\it CH$_2$NH} & 2.2 $\pm$ 0.2 &  3.4 $\pm$ 0.4 & 3.2 & $<$69 & $\dots$  & 160  \\ 
{\it HOC$^+$} & 1.1 $\pm$ 0.2 &  1.3 $\pm$ 0.3 & 1.2 & $<0.2$ & $\dots$  & 140  \\ 

\\[0.1mm]
{\bf Zw049.057}\\
{\it HCN 3--2} & 18.8 $\pm$ 0.4 &  64.4$\pm$ 1.5 & 60.7 & $172 \times 137$   & 25 & $\dots$ \\
{\it HCN-VIB 3--2} & 4.2$\pm$ 0.2 & 6.8 $\pm$ 0.5& 6.4 & $<$86 & $\dots$  & 250 \\
{\it HCO$^+$ 3--2} & 7.3 $\pm$ 0.2 &  25.3 $\pm$ 0.7 & 23.8 &  $133 \times 94$  & 30 & $\dots$   \\ 
{\it CH$_2$NH} & 4.6 $\pm$ 0.3 &  11.3 $\pm$ 1.1 & 10.8 & $154 \times 80$ &  $\dots$ & 250  \\ 
{\it HOC$^+$} &1.7 $\pm$ 0.2 &  2.6 $\pm$ 0.4 & 2.5 & $<0.4$ & $\dots$  & 200  \\ 

\\[0.1mm]
{\bf Arp220W}\\
{\it HCN 4--3} & 130 $\pm 5$  & 294 $\pm 15$  &  670 &  $310 \times 250$ & 146 & $\dots$    \\
{\it HCN-VIB 4--3} & 55 $\pm$ 3 &  70 $\pm 5$ &  160 & $62 \times 50$ & 133 & 350$^{\star}$ \\
{\it HCO$^+$ 4--3} & 55 $\pm$ 2  & 123 $\pm 7$  &  286 & $310 \times 250$ &140 & $\dots$    \\ 

\\[0.1mm]
{\bf IRAS17208}\\
{\it HCN 4--3} & 53.7 $\pm$ 0.1 &  86.0 $\pm$ 0.2 & 941 & $307 \times 256$ & 95 & $\dots$    \\
{\it HCN-VIB 4--3} & 8.0 $\pm$ 0.2 &  8.0 $\pm$ 0.2 & 87 & $<$136 & $\dots$  & 450$^{\star}$\\
{\it HCO$^+$ 4--3} & 30.8 $\pm$ 0.05 &  55.8 $\pm$ 0.05 & 610 &  $350 \times 299$  & 100 & $\dots$  \\ 
{\it H$_2$S} & 36.3 $\pm$ 0.1 &  45.3 $\pm$ 0.2 & 496 & $196 \times 136$ & 95 & 450 \\ 

\\
\hline \\
\end{tabular} 

\begin{minipage}[h]{0.8\hsize}
$^{\dag}$Calculated from two-dimensional Gaussian fits to the integrated intensity emission  using the AIPS task IMFIT which deconvolves the clean beam
from the fitted component size. The accuracy and errors of this process are discussed in \citet{condon97}.   \\
$^{\star}$From Gaussian fits where the line is blended with HCO$^+$.
Limits to HCN-VIB brightness temperatures are $T_{\rm B}$(HCN-VIB)$>$20 K for IC860, $>$6 K for Zw049.057, $>$8 K for IRAS17208-0014 and 68 K for Arp220W. \\
For IC860 and Zw049.057 1\asec=286 pc, for Arp220W 1\asec=388 pc  and for IRAS17208-0014 1\asec=853 pc.  

\end{minipage}

\end{table*}


\begin{table}[tbh]
\caption{\label{t:cont} Continuum fluxes and source sizes$^{\dag}$.}
\begin{tabular}{lcccc}
\hline
\hline
\\[0.1mm]
Name & Peak & Integrated & $\theta$ & $T_{\rm B}$\\
          & [mJy] & [mJy]  & [pc]  & [K]\\ 
\hline
\\[0.1mm]
IC 860   &  $35 \pm 2$ & $35 \pm 2$ & $<$37  &  $>$43\\
Zw 049.057 & $20\pm 1$ & $44\pm 3$ & $145 \times 86$ & $>$5 \\
Arp220W & $283 \pm 3$ & $384 \pm 4$ & $62 \times 50$ &  $>$160 \\
IRAS17208 & $87 \pm 1$ & $116 \pm 2$ & $248 \times 162$ & $>$40\\
\hline
\end{tabular} 
\\
\newline
\\
\begin{minipage}[h]{0.9\hsize}
$^{\dag}$Continuum levels were determined through a zeroth-order fit to line-free channels in the uv-plane.
Source sizes (diameters) are full width half maximum (FWHM) two-dimensional Gaussian fits. 
The continuum fluxes and source sizes are determined for 260 GHz for IC860 and Zw049.057 and for 350 GHz for IRAS17208-0014 and Arp220. The
continuum values for Arp220W are from Mart\'in et al. (in prep.) and are consistent with those reported in \citet{sakamoto08}.
\end{minipage}
\end{table}


\section{Discussion}

\subsection{Self-absorption in the HCN and HCO$^+$ lines}
\label{s:self2}

The nuclear HCN and HCO$^+$ spectra are double peaked in contrast to the HCN-VIB lines (and also the CH$_2$NH, H$_2$S and HOC$^+$ lines).  We argue that this is
due to a combination of {\it self- and continuum absorption} in the HCN and HCO$^+$ lines where photons at the line centre are reabsorbed by cooler gas 
 in the external parts of the HCN-VIB emitting core and plausibly also in more extended regions.
(see Fig.~\ref{f:cartoon} for an illustration). 
A temperature gradient and high column densities will therefore result in self-absorbed lines as seen also in embedded sources in the Milky Way \citep{rolffs11a}. 
The self-absorption obliterates any sign of the hot dense nucleus in the HCN and HCO$^+$ lines impeding their use as tracers of the properties of extremely dust-embedded nuclei. 
The impact of self-absorption is further discussed in Sec.~\ref{s:impact} below.

\medskip
\noindent
How do we know that the twin-peaked HCN and HCO$^+$ lines are not caused by a real absence of dense molecular gas in the nucleus? 
We need other lines to compare with to determine the cause of the absorption. 

\subsubsection{Evidence from the HCN-VIB line}

We can test the self-absorption theory for HCN through comparing the intensity in the vibrational $\nu_2$ state (both the 1$e$ and 1$f$ line) to that of the vibrational ground state
 - $I_u/I_l$ in eq.~\ref{e:vib}). Unless the HCN-VIB line is masing, $I_u < I_l$. The $I_u$=$I$(HCN-VIB)(1$e$+1$f$) and for thermal excitation the 1$e$ line is as bright as the
1$f$ line and hence $I_u$=$2\times I$(HCN-VIB). Thus, for $I_u < I_l$ to hold, the {\it HCN-VIB/HCN line intensity ratio, ${\cal R}_{\rm vib}$,} should be $<$0.5. 
 Note, that here we are assuming that all of the HCN emission is emerging from the same gas that produces the HCN-VIB line.  Any additional HCN-emission would further lower the "allowed"
${\cal R}_{\rm vib}$ ratio.

\smallskip
\noindent
We can now inspect the nuclear spectra  and determine ${\cal R}_{\rm vib}$ on a channel-by-channel basis.
For {\it IC860 and Arp220W} ${\cal R}_{\rm vib}$ exceeds unity close to systemic velocities and therefore line photons of both ground state HCN and 1$e$ HCN-VIB, which appears close to
the HCN $\nu$=0 line centre (see Sect.~\ref{s:intro2}), are being absorbed (see zoomed-in example for IC860 in Fig.~\ref{f:velo}).
However for {\it Zw049.057} the ratio is ${\cal R}_{\rm vib}$=0.5 and for {\it IRAS17208-0014} ${\cal R}_{\rm vib}$=0.25 and additional evidence is required to claim that the HCN and HCO$^+$ line shapes are due to self-absorption.   Furthermore, continuum absorption by the optically thick HCN and HCO$^+$ lines
can be important, given that the continuum brightness is similar to the continuum-subtracted line brightness at central velocities.

\subsubsection{Comparing with other lines}

We can use other, optically thin, species to trace the dense gas in the nuclear region. In Appendix~\ref{s:A1} we discuss the CH$_2$NH line emission in Zw049.057 and the
H$_2$S line emission of I17208-0014 and how they support the notion of self-absorbed HCN and HCO$^+$ in these galaxies.  In Appendix~\ref{s:A2} we also discuss how the
HCO$^+$/HOC$^+$ ratio is also consistent with self-absorbed HCO$^+$.

\subsection{Impact of self-absorbed HCN and HCO$^+$ lines in galaxy nuclei}
\label{s:impact}

The effect of self absorption on the HCN and HCO$^+$ lines (see illustration in Fig.~\ref{f:cartoon}) demonstrates that the emission cannot be viewed as emerging
from an ensemble of clouds that are independent of each other - but that emission from clouds closer to the nucleus is being absorbed by clouds farther out.
The notion that the spectrum observed towards a galaxy involves "counting clouds" - that may be individually optically thick but collectively "optically thin"
(i.e., not overlapping each other in the 2D (sky plane) or the velocity (line-of-sight) space) does not hold for these galaxy nuclei.

This breaking down of the {\it ensemble approximation} has been predicted by \citet{downes93} who suggested that (U)LIRG nuclei should be viewed as a single cloud. 
The ensemble approximation is appropriate when the mean gas surface density is low or moderate, but becomes less and less valid when
observing a dense ULIRG/compact nucleus at increasing resolution. A similar conclusion that emission is not emerging from individual, non-overlapping clouds 
in the nuclei of Arp220 was also drawn by \citet{scoville15}.

\medskip
\noindent
When lines of HCN and HCO$^+$ become self-absorbed they cannot be reliably used to probe the central regions of galaxies on scales where the absorption occurs - for our
sample galaxies this means $r<$100 pc.  For the studied galaxies this is also where most of the emission comes from, in contrast to many nearby galaxies with central molecular
zones extended over larger regions. Thus, the absorbed HCN and HCO$^+$ line profiles are strongly impacting our understanding also of the global properties of these galaxies.
Very recently self-absorbed CO 6--5 has been detected towards both nuclei of Arp220 by \citet{rangwala15}.



\subsection{HCN-VIB lines as probes of extreme mid-IR surface brightness and luminosity}
\label{s:buried}

Vibrationally excited lines of HCN are excellent probes of the nuclear dynamics and serve as proxies for the mid-IR luminosity density in deeply obscured nuclei
where $N$(H$_2$)$>10^{24}$ $\cmmt$ and where dust temperatures exceed 100 K.
Mid-IR excitation of HCN is favoured in regions of high dust temperature and with significant gas and dust column densities. This is because an 
$T_{\rm B}$(14$\mu$m)$\gapprox$100 K is necessary to start to excite the HCN vibrational ladder and this requires an optically thick mid-IR source - which
in turn requires an H$_2$ column density $N$(H$_2$) $ \gapprox 2 \times 10^{23}$ $\cmmt$ (see Appendix~\ref{s:B} for a discussion). Note that here we assume that
the molecules receive the exciting radiation isotropically. If they do not then the background source must be even hotter to make up for the loss in solid angle.

\medskip
\noindent
The HCN-VIB lines of the sample galaxies are very luminous (see discussion in Sect~\ref{s:luminosity}) and lower limits to $T_{\rm B}$(HCN-VIB) exceed 20 K for IC860
and Arp220W (see footnote to Tab.~\ref{t:flux}). Therefore,  the filling factor of hot dust and gas is extremely high in the inner tens of pc of the sample galaxies - much
higher than on similar size scales in Milky Way massive star forming regions or in extended starbursts. The implication is that the HCN-VIB emission we
detect towards these galaxies cannot be emerging from a collection of Milky-Way-like hot cores. The high filling factor of hot, optically thick gas and dust suggests that we
can use the minimum 14$\mu$m brightness temperature of 100 K to estimate a {\it lower limit} to the mid-infared luminosity.

\smallskip
\noindent
We assume a nuclear spherical "cloud" of dust for which we adopt a simple blackbody where $L$=4$\pi r^2 \sigma T^4$ ($\sigma$ is Stefan Boltzmann's constant).
Taking $r$ from Tab.~\ref{t:flux} (i.e. $\theta$/2) and for $T_{\rm B}$(14$\mu$m)=100 K the corresponding $L$(14$\mu$m) is $6 \times 10^{10}$ \lsun\ for IC860, $3 \times 10^{10}$ \lsun\ for Zw049,  $2 \times 10^{11}$ \lsun\ for Arp220W and $8 \times 10^{11}$ \lsun\ for IRAS17208.  In terms of temperature this is a lower limit since at 100 K the
excitation of the HCN-VIB line will not be very efficient.  We know from the self-absorption structure of HCN and HCO$^+$ (see Sect.~\ref{s:self2}) that there is a  temperature gradient and a significant fraction of the HCN-VIB line will emerge from regions of higher temperatures even closer to the nucleus. Thus, when the radius of
the emitting region is known - like it is for Arp220W - this value will be a firm lower limit to the real luminosity.  For Arp220W,  high resolution observations the submm continuum
suggest an optically thick, peak dust temperature of 160- 200 K \citep{downes07,sakamoto08,matsushita09,wilson14} demonstrating that an assumption of $T_{\rm B}$(14$\mu$m) of 100 K is too low.
We therefore conclude that these compact mid-IR nuclei may constitute 20\% - 100\% of the total observed $L$(IR). 

\smallskip
\noindent
From a lower limit to the $T_{\rm B}$(14$\mu$m) we can estimate an intrinsic (unobscured) 14$\mu$m surface brightness, $\Sigma_{14}$. With the assumption of blackbody radiation
$$ 
\Sigma_{14}={L \over {2A}}=2\pi r^2 \sigma T^4/\pi r^2=2\sigma T^4. 
$$
For $T_{\rm B}$(14$\mu$m)=100 K,  $\Sigma_{14}$=$3 \times 10^{13}$ \lsun\ kpc$^{-2}$, for a $T_{\rm B}$(14$\mu$m) of 150 K $\Sigma_{14}$=$1.5 \times 10^{14}$ \lsun\ kpc$^{-2}$ and for 200 K, $\Sigma_{14}$=$5 \times 10^{14}$ \lsun\ kpc$^{-2}$.  

\noindent
Note again, that a more sophisticated approach to the radiative transport and source structure is necessary for a more accurate discussion of the surface brightness and 
mid-IR luminosity.

\subsubsection{Attenuated dust SEDs}

The luminous mid-IR nucleus, which is exciting the HCN-VIB line, is hidden inside massive layers of dust and its intense emission may become absorbed and re-emitted at longer
wavelengths – altering the dust spectral energy distribution (SED). The unattenuated SED may be hot (e.g. AGN-like) while the observed SED instead appears cooler and more like
that of a typical starburst.  

\begin{figure}
\includegraphics[scale=0.33]{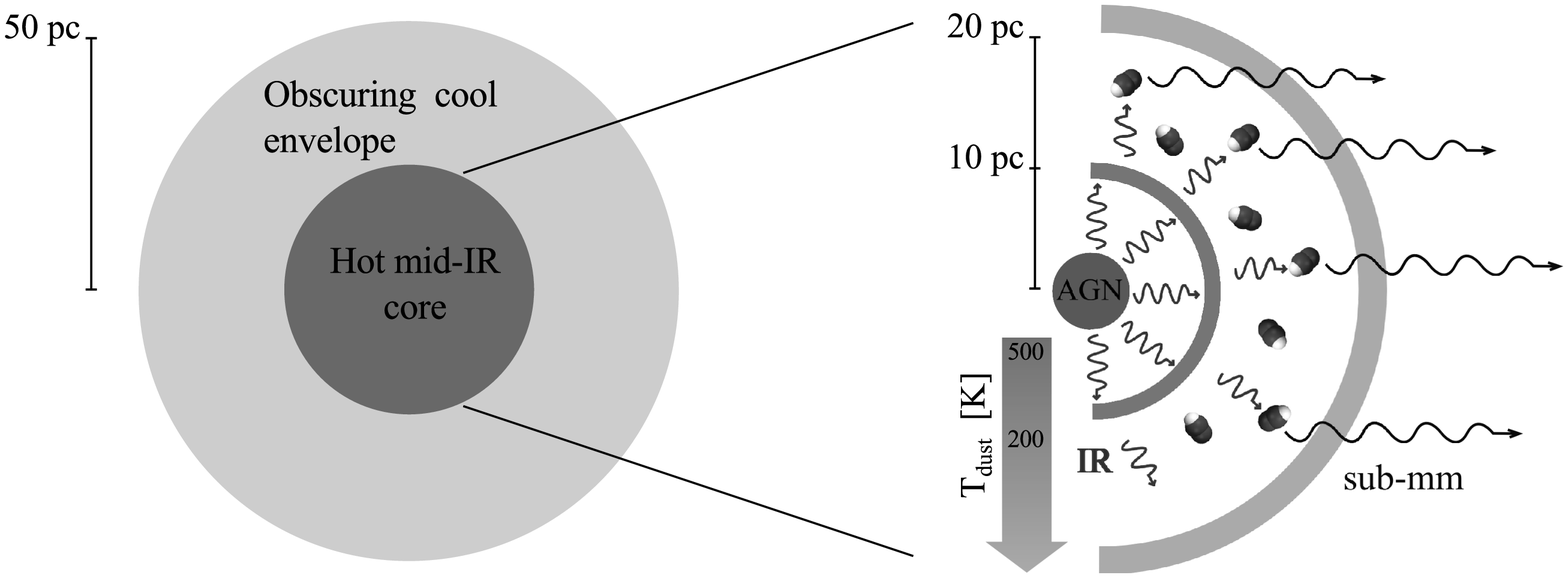} \\[6 mm]
\includegraphics[scale=0.70]{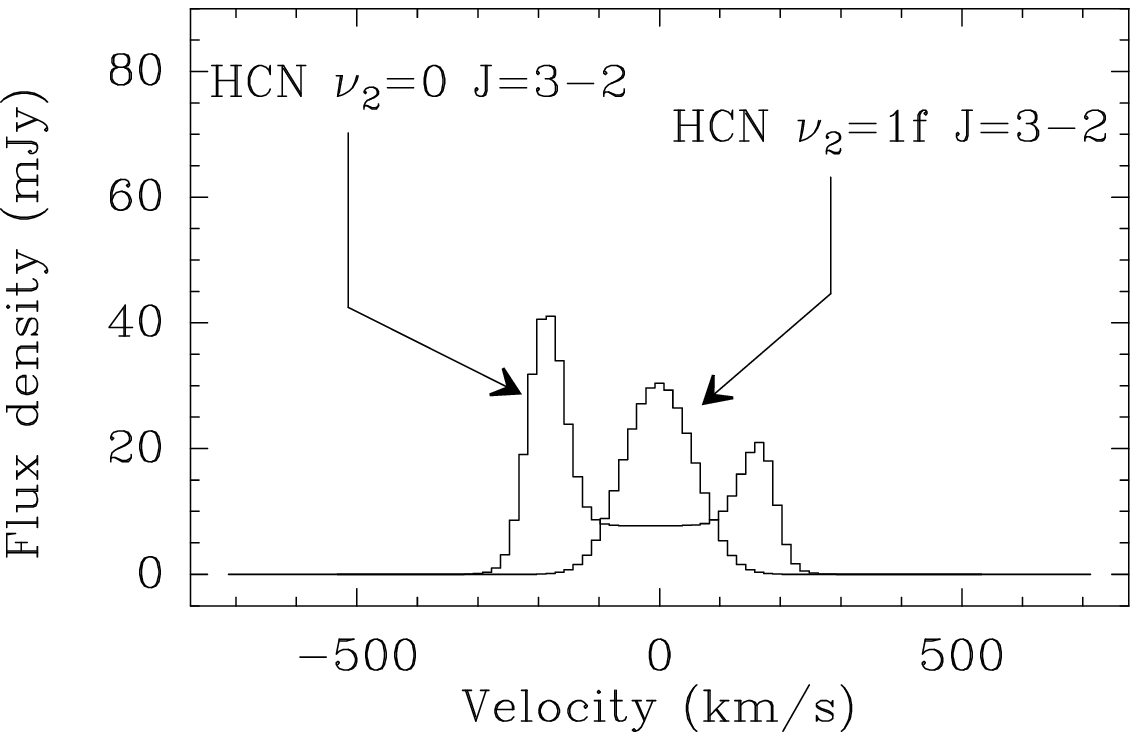} \\[6 mm]
\includegraphics[scale=0.75]{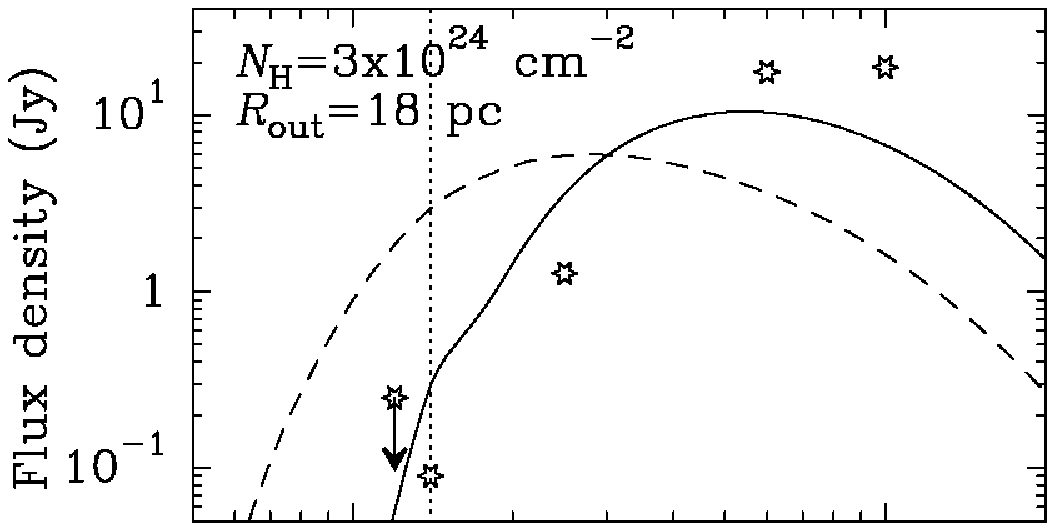}
\caption{\label{f:cartoon} Upper panel: Illustration of the vibrationally excited (HCN-VIB) line emission emerging from a buried mid-IR core -surrounded by cooler
material giving rise to the self-absorption feature in the HCN line.  An embedded AGN nucleus would be one possible scenario that is powering the bright HCN-VIB
and causing the temperature gradient with the resulting self-absorbed HCN and HCO$^+$ lines. Middle panel: Illustration of a possible scenario that may account
for the observed HCN and HCN-VIB emission in IC860 with a buried  compact heating source,  
where the assymmetric HCN spectrum is caused by an inflow of gas. The model was generated with the tool described in \citet{gonzalez99}, which includes line overlaps \citep{gonzalez97}. Note that these examples assume a spherical geometry while our results may imply a more torus- or disk-like morphology. 
Lower panel: This is an illustration of how the dust spectral energy distribution (SED) for IC860 is shifted from shorter to longer wavelengths by the absorption of the emission from the hot buried source. The dashed curve is the intrinsic unobscured SED and the stars and the solid line indicate the observed flux and the fitted SED.}
\end{figure}

In the conservative (100 K) estimate of $L$(14$\mu$m) above for IC860 the unattenuated 14$\mu$m continuum flux from the buried source should be $\approx$2.5 Jy. 
However,  the observed flux density is more than a factor 30 lower  \citep{lahuis07}. In Fig.~\ref{f:cartoon} (lower panel) we show an example of a possible intrinsic SED of
IC860 together with the observed SED. Another example is IRAS17208-0014 where our HCN-VIB observations imply the presence of a $\gapprox$4Jy compact ~mid-IR source while
high resolution imaging at mid-IR wavelengths only finds an extended disk of a total flux of 0.2 Jy \citep{soifer00} . The authors note the striking discrepancy between the 2" mid-IR
source and the $0."26 \times 0."32$ 8.6 GHz radio source which sets IRAS17208-0014 aside from the other ULIRGs they studied. The bright HCN-VIB line towards the nucleus of IRAS17208-0014
suggests that the solution to this discrepancy is an extreme obscuration of the nucleus at mid-infrared wavelengths.

The trapped radiation from an embedded source may also aid in raising the internal temperature significantly. High column densities in all directions lead to the dust being
optically thick in the infrared and this radiation is unable to escape and diffuses outwards by multiple absorption/emission events until the dust is optically thin to its own radiation. This process is discussed in \citet{rolffs11b} where this "greenhouse effect" (Gonz\'alez-Alfonso et al, in prep) is used to explain the high masses of hot gas near an embedded protostar.



\begin{figure}[h]
\includegraphics[scale=0.5]{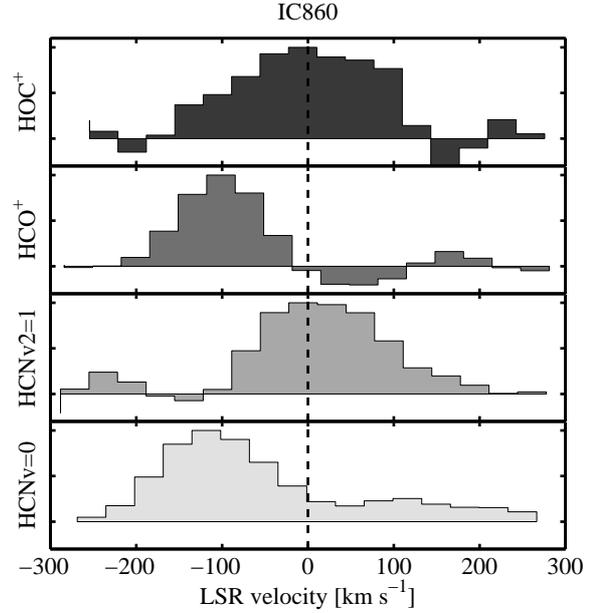}
\caption{\label{f:velo} Zoom in on the spectral structure of  HOC$^+$, HCO$^+$ (top two panels) and  HCN-VIB,HCN (lower two panels) of IC860 showing the self-absorbed line
shapes of HCN, HCO$^+$ while the other two lines peak at systemic velocity. The intensities have been normalized.
}
\end{figure}

\subsection{The nature of the buried source}
\label{s:buried}

The HCN-VIB lines are emerging from highly compact ($r<$ 17-75 pc) regions centered on the nucleus of the galaxy. The sphere of influence \footnote{The region where its gravitational potential overwhelms that of the galaxy with $r_{\rm g}=2GM_{\rm BH}/{\sigma_{\rm S}}^2$  ($\sigma_{\rm S}$ is the stellar velocity dispersion). It is usually larger
than the region inside the Bondi accretion radius $r_{\rm acc}=2GM_{\rm BH}/{c_{\rm S}}^2$ ($c_{\rm S}$ is the speed of sound in the ISM) where the gravitational energy of the SMBH dominates over the thermal energy of the gas.}
of the SMBH in the core of the galaxy is typically 10-20 pc.
Therefore, the HCN-VIB emission is emerging near
to this region and thus originates in a transition-zone constituting an possible evolutionary link between the larger scale host galaxy and the pc-scale environment near the SMBH.

\medskip
\noindent
A {\it dust-enshrouded AGN} would be consistent with the high mid-IR surface brightness and a temperature gradient causing the HCN and HCO$^+$ self-absorption. The SMBH of IC860
(for example) would be accreting at 12 - 30\% of its Eddington limit\footnote{The K-band luminosity of IC860 \citep{brown14} implies a mass of the SMBH of $\approx 2 \times 10^7$ \msun\ (using the prescription described in section 3.4 of \citet{sakamoto13}). An SMBH accreting at its Eddington limit would have $L=L_{\rm Edd} > 10^{4.42}$ $M$/\msun \lsun.} to be the cause of the excitation of the HCN-VIB line and 50-100\% of the FIR luminosity.
An alternative possibility is that the intense mid-IR emission emerges from a compact starburst. The mid-IR surface brightness of  $>5 \times10^{13}$ \lsun\ kpc$^{-2}$ exceeds the limiting Eddington flux of a cool, optically thick starburst. However, a {\it hot, optically thick starburst} mode with a bulk temperature $T>$200 K may exist according to \citet{andrews11} and they propose that its Eddington flux is $F_{\rm Edd} \propto 10^{15}\, L_{\odot}\, {\rm kpc}^{-2} ({\Sigma_{g} \over {10^6\, M_{\odot}\, {\rm pc}^{-2}}})$ (from their equation (7) where $\Sigma_g$ is the gas surface density). 

For both cases the HCN-VIB line is tracing a {\it phase of exceptional growth} in the nucleus of the galaxy in the form of a rapidly accreting SMBH and/or the fast build-up of a nuclear stellar spheriod/disk.  


\subsubsection{A rotating dust torus?}

Interestingly, the HCN-VIB emission of IC860 and Zw049.057 also shows a faint signature of a twin velocity peak in the pV diagram (see Fig.~\ref{f:pv}). This implies that the emission is not arising from a sphere or filled disk but is consistent with a torus-like structure. However, small (pc-scale) optical depth effects in an opaque dust core cannot be excluded, or the possibility that also the HCN-VIB line is partially self-absorbed.
The width of the HCN-VIB line suggests that the enclosed mass inside $r$=18 pc exceeds $5 \times 10^7$ \msun\ for IC860. The K-band luminosity implies a mass of the SMBH of 
$2 \times 10^7$ \msun\ (see Sect.~\ref{s:buried}) which has error bars of roughly a factor of five. Higher resolution observations of the vibrationally excited HCN line will allow determinations of the SMBH mass of higher accuracy.


\subsection{Where do we find HCN-VIB emission?}

Even though there has been no systematic survey, the galaxies with HCN-VIB emission detections so far are either ULIRGs or LIRG early type spirals (see Tab~\ref{t:ratio}). 

The detection of HCN-VIB emission requires the presence of high masses of gas and dust and a compact heating source.  Early type spirals have steeper
potential wells than late type spirals which may more easily facilitate the formation of compact dust cores. However, optically thick,  dusty nuclei should be easier to find in early
type spirals than in late type ones since the former generally do not have contaminating star formation in their disks. 
A survey will reveal whether the HCN-VIB lines preference for early type spirals is real or a result of observational bias, but it is interesting to note
that \citet{mignoli13} find that moderate luminosity, obscured type-2 AGNs occur primarily in early-type Sa-Sb galaxies.

The link between major mergers and ULIRGs is well known \citep{,sanders96,bekki99,elbaz03} and luminous, dense circumnuclear disks of gas and stars are a natural consequence of gas-rich galaxy mergers since they trigger massive inflows of gas \citep{barnes02, mayer07}.  In the very inner region the gas may contribute to the rapid growth of SMBHs and/or the build up of nuclear stellar populations. The incidence of HCN-VIB line emission in ULIRG major mergers will give important clues to when during the ULIRG phase the nuclear region is undergoing accelerated growth.

\subsubsection{The $L$(HCN-VIB)/$L$(FIR) ratio}
\label{s:luminosity}

In Tab.~\ref{t:ratio} we list the $L$(HCN-VIB)/$L$(FIR) ratio for the eight galaxies that have HCN-VIB detections so far: four lenticulars (NGC1377, NGC4418, IC860 and Zw049.057)
and four ULIRGs (IRAS20551-4250, Arp220, Mrk231, IRAS17208-0014). We have also added the Galactic massive star formation region SgrB2(M) to illustrate that the ratios for most of
the HCN-VIB detected galaxies are higher by an order of magnitude than those of a forming stellar cluster in the centre of the Milky Way. Even if we populated the inner
region of (for example) IC860 solely with SgrB2(M)-like regions this would not reproduce our observations. We have also added
the ULIRG IRAS13120-5453 to Table~\ref{t:ratio} since we have high signal-to-noise ALMA observations (Privon et al in prep.) with a good upper limit to the HCN-VIB line. 

\medskip
\noindent
There is no obvious relation between $L$(HCN-VIB)/$L$(FIR) and FIR luminosity. The ratio will naturally be dependent on the temperature of the exciting dust source and
the fraction of the $L$(FIR)  emission that is emerging from the same region as the HCN-VIB emission. However, a trend to be explored further is the correlation with the 
strength of molecular outflows  (see Table~\ref{t:ratio}). The rapidly evolving nature of the studied nuclei is illustrated by what seems to be in- and outflow signatures in the self-absorbed HCN and HCO$^+$ lines, but also in other absorption- and emission lines. {\it All galaxies with a detected HCN-VIB line show evidence of non-circular motions: inflows, outflows or both.}
Tentatively, the galaxies with the relatively brightest HCN-VIB lines show signs of slow non-circular motions: NGC4418, Arp220W, IC860, Zw049.057, IRAS17208-0014 (see Tab.~\ref{t:ratio} for references). (The reversed P-Cygni profiles we find for IC860 could be evidence of infall, but more observations and modelling are required to determine their precise nature.)
Galaxies with fast outflows (i.e. where the outflow velocities exceed the galaxy's rotational (virial) velocity) have fainter HCN-VIB signatures. The ULIRGs Mrk231, IRAS20551 and
IRAS13120 have prominent molecular outflows, seen in the OH 119 $\mu$m line, with maximum velocities $>$1000 \kms (see Tab.~\ref{t:ratio} for references). The lenticular galaxy NGC1377 has a prominent molecular outflow in relation to its relatively modest escape velocity and mass \citep{aalto12b}.

Apparently in contrast, \citet{spoon13} find, using the 119 $\mu$m OH absorption line, that the highest outflow velocities are found in galaxies with deep mid-IR silicate absorption
indicating that they are embedded.  It is possible that the HCN-VIB line is probing a deeper, more extreme form of nuclear obscuration - and possibly also more transient - than
the one traced by the mid-IR  silicate absorption.  Indeed, \citet{roche15} find for NGC4418 that a large fraction of the silicate absorption is arising from "cool material outside the warm
molecular core". In addition, \citet{gonzalez15} find that the silicate absorption is biased towards relatively unabsorbed mid-IR emitting regions and propose high-lying OH absorption lines
as tracers of embeeded, hot nuclei. (The correlation between HCN-VIB and these highly excited OH lines will be explored in future work.)

A potential scenario  is therefore that the HCN-VIB phase peaks {\it right before} the nuclear feedback mechanism is fully developed. This would
be consistent with our conclusion in Sect.~\ref{s:buried} that the HCN-VIB line is probing a transient phase of rapid nuclear growth. A possible complication to this picture is the
recent discovery of a 800 \kms\ CO 2-1 outflow in the HCN-VIB luminous ULIRG IRAS17208-0014  \citep{garcia15}. No fast 119 $\mu$m OH wind has been found
and it is possible that these absorption studies  may miss in- and outflows that occur deep inside an optically thick dust shell.

\begin{table}[tbh]
\caption{\label{t:ratio} $L$(HCN-VIB) vs $L$(FIR)}
\begin{tabular}{lcccc}
\hline
\hline\\[0.01 pt]
Name & log $L$(FIR) & ${L({\rm HCN-VIB}) \over L({\rm FIR})}$ & Molecular& Ref \\[1.0 mm]
    & & [$10^{-8}$] & Outflow$^{\dag}$ \\ 
\hline
\\[0.1mm]
$J$=3--2 \\
NGC4418 & 11.08 & 3.8 & n$^{\star}$ & [1] \\
IC 860   &  11.17 &  3.2 & n$^{\star}$ & This work \\
Zw 049.057 & 11.27 & 3.4 & y$^{\star}$ & This work  \\
Mrk231 & 12.37 & 0.5 & Y & [2] \\
\hline \\[-1.8 mm]
{\it SgrB2(M)} & 6.8 & 0.3 &  y &  [3] \\

\\
$J$= 4--3 \\
NGC1377 & 10.13 & 1.0 & Y & [5] \\
NGC4418 & 11.08 & 9.5 & n$^{\star}$ & [1] \\
IRAS20551 & 12.00 & 0.4 & Y &[4] \\
Arp220W & 12.00 & 15 & y & This work \\
IRAS13120 & 12.26 & $<$0.2 & Y & [6] \\
IRAS17208 & 12.39 & 3.6  & y &This work\\
\hline
\end{tabular} 
\\
\newline
\\
\vspace{2mm}
\begin{minipage}[h]{0.95\hsize}
[1] \citet{sakamoto10,sakamoto13}, [2] \citet{aalto15}, [3] \citet{rolffs11a}, [4] \citet{imanishi13}, [5] Aalto et al (in prep.), [6] Privon et al. (in prep.)\\
$^{\dag}$A capital Y indicates molecular outflow velocities exceeding the escape velocity for the nuclear region of the galaxy. References for the outflows/infall: 
NGC4418 \citep{gonzalez12, sakamoto13, costagliola13, veilleux13}; IC869 - this work; Zw049.057\citep{falstad15});
IRAS17208-0014 \citep{veilleux13} (but note recent work by \citet{garcia15} where CO 2--1 outflow velocities of 800 \kms are found), 
Mrk231 \citep{fischer10,veilleux13, gonzalez14}, IRAS13120-5453 and IRAS20551 \citep{veilleux13},  Arp220W \citep{sakamoto09,veilleux13},
SgrB2(m) \citep{rolffs11a}.\\
$^{\star}$Minor axis dust lanes in the form of cones suggesting current or past outflows have been found in NGC4418 \citep{sakamoto13} and also in IC860 (unpublished
Nordic Optical Telescope U,B images by F. Costagliola). Also in Zw049.057 a peculiar minor axis outflow-like dust feature has been found \citet{scoville00}.

\end{minipage}
\end{table}

\subsubsection{HCN-VIB as a survey tool for obscured high-surface brightness activity}
\label{s:survey}

Our observations show that luminous extragalactic HCN-VIB emission is emerging from compact nuclei of high mid-IR surface brightness and that vibrationally excited HCN is a strong spectral signature of a nucleus in rapid growth - either through Eddington limited (or near-Eddington) nuclear starburst activity or of accreting AGNs. 
The HCN-VIB line is an effective survey tool to find high surface brightness nuclei hidden behind high masses of cooler dust and gas - even with observations at low spatial resolution.
Detecting the vibrationally excited HCN also allows us to reconstruct the intrinsic, hotter SED of the luminosity source. 

The HCN-VIB line from a nuclear compact core will show up even if there is more extended star formation which masks and confuses other tracers such as the  [C II]/FIR ratio and radio
continuum imaging.  Galaxies with HCN-VIB emission therefore constitute a sample where to search for deeply obscured AGN activity. Any X-ray emission should be severely attenuated - even hard($>$ 5~keV) X-ray emission if column densities exceed $N$(H$_2$)$>10^{24}$ $\cmmt$. Follow-up, high resolution HCN-VIB and continuum studies will help measure the luminosity density and enclosed mass which will put strong constraints on the nature of the buried source. Thus, studies of vibrationally excited lines can be used to reveal and probe a population of deeply dust-buried AGNs promising to shed light on how the SMBH growth connects to that of the stellar bulge and to enable studies of the competition between star formation and accretion.
Emerging new techniques utilizing He$^+$ recombination lines may further assist in separating accreting SMBHs from hot compact starbursts \citep{scoville13}.

\section{Conclusions}

\begin{itemize}

\item High resolution (0."4) IRAM PdBI and ALMA mm and submm observations of the LIRGs IC860 and Zw049.057, and the ULIRGs Arp220 and IRAS17208-0014 reveal luminous vibrationally excited HCN $J$=3--2 and 4--3 $v_2$=1f (HCN-VIB) line emission emerging from buried, compact ($r<17-70$ pc) nuclei. They have very high inferred mid-infrared surface brightness $>5 \times 10^7$ \lsun pc$^{-2}$ and are powered by dust enshrouded accreting SMBHs and/or hot ($>200$ K) extreme starbursts.  The HCN-VIB emission may be the
signpost of an enshrouded nucleus undergoing rapid evolution either through the growth of its SMBH or a compact stellar spheroid. In either case, the HCN-VIB emission is emerging from inside the inner few tens of pc which consitutes an important transition region and a potential evolutionary link between the larger scale host galaxy disk and the pc-scale environment
near a SMBH. 

\item In contrast, we show evidence that  the ground vibrational state rotational lines of  HCN and HCO$^+$ $J$=3--2 and 4--3 (standard tracers of dense gas in galaxies) fail to probe the highly enshrouded, compact nuclear regions due to strong effects of self-absorption (caused by a temperature gradient and large line-of-sight column densities) and absorption of continuum. We advocate HCN-VIB lines as better tracers of the dynamics, mass and physical conditions of (U)LIRG nuclei when H$_2$ column densities exceed $10^{24}$ $\cmmt$ and when there is a buried compact luminosity source. The vibrationally excited HCN allows us to identify galaxies where nuclear mid-infrared emission may be strongly absorbed and re-emitted at longer wavelengths - and to 
reconstruct the intrinsic, hotter dust spectral energy distribution (SED) of the buried luminosity source. 

\item With this study HCN-VIB lines have now been found in eight (U)LIRGs with AGN signatures and/or evidence of compact mm/submm continuum emission. They are either ULIRG mergers
or early-type spirals with centrally concentrated molecular gas and dust.  The embedded galaxy nuclei show strong signatures of non circular motions that could be interpreted as infall and/or molecular outflows - also in the self-absorbed HCN and HCO$^+$ spectra. IC860 have reversed P-Cygni profiles that may be evidence of inflowing gas.
A tentative conclusion is that galaxies without evidence of fast outflows (i.e. velocities greater than rotational)  have the strongest HCN-VIB line in relation to their IR luminosities. This implies that 
the HCN-VIB emission peaks before the onset of strong feedback.

\item All galaxies with detected HCN-VIB lines show very rich molecular spectra and we report the detection of luminous emission lines of CH$_2$NH and HOC$^+$ and vibrationally excited HC$_3$N in IC860 and Zw049.057 also lines of CH$_3$OH are detected.  In IRAS17208-0014, very bright H$_2$S emission is found.

\end{itemize}

\begin{acknowledgements}
We thank the IRAM PdBI staff and the Nordic ALMA ARC node for excellent support.  SA acknowledges partial support from the Swedish National Science Council grant 621-2011-4143. 
KS was supported by the MOST grant 102- 2119-M-001-011-MY3. This paper makes use of the following ALMA data: ADS/JAO.ALMA2012.1.00453.S and  2012.1.00817.S. 

\end{acknowledgements}

\bibliographystyle{aa}
\bibliography{vib_ref}

\begin{appendix}

\section{Self-absorbed HCN and HCO$^+$ lines: Comparing with other species}
\label{s:A1}

\subsection{Comparing with the H$_2$S and CH$_2$NH lines}
\label{s:A1}
Both the CH$_2$NH and H$_2$S molecules have high dipole moments  ($\mu_{\rm a}$=1.340 D
and $\mu$=0.97 D respectively) and the transitions observed here have critical densities in excess of 10$^7$ $\cmmd$ \citep{dickens97,crockett14}. The CH$_2$NH/HCN abundance
ratio is typically 10$^{-3}-10^{-2}$ \citep{dickens97} suggesting that the CH$_2$NH line emission should suffer self-absorption only in extreme conditions. We find that the CH$_2$NH
line in Zw049.057 has a single peak at systemic velocity and the emission has a maximum at the same spatial position as the HCN-VIB line. Similarly, in IRAS17208-0014 the 369 GHz
H$_2$S line clearly peaks at systemic velocities and $I$(H$_2$S)$>$$I$(HCN) between 0 and -200 \kms. Even though it is possible for the abundances of the two molecules to be similar
in the inner regions of hot cores \citep{crockett14} the H$_2$S abundance must be two orders of magnitude greater than that of HCN for $I$(H$_2$S)$>$$I$(HCN). This would constitute an hitherto unknown astrochemical scenario. Thus the simplest explanation is that HCN is self-absorbed in the inner region of Zw049.057 and IRAS17208-0014.

\subsection{Comparing with the HOC$^+$ line}
\label{s:A2}
Zw049.057 and  IC860 show intensely bright  HOC$^+$ 3--2 line emission with $I$(HOC$^+$)$>$$I$(HCO$^+$) at
systemic velocity (see example for IC860 in Fig.~\ref{f:velo}).  Assuming the excitation of the two molecules is similar, this either requires that HOC$^+$ abundances exceed those of HCO$^+$ - or that HCO$^+$ is self absorbed.
The abundance ratio between the two molecules is a measure of the electron abundance and in Galactic dense clouds HCO$^+$/HOC$^+$ is typically a few times $10^3$. In more extreme
regions such as the photon dominated regions (PDRs) of the starburst galaxy M82 and near the nucleus of the Seyfert galaxy NGC1068, this ratio is found to range between 44 and 1  \citep{fuente08,usero04}.  In no source has HOC$^+$ yet been found to be more abundant than HCO$^+$ and chemical models do not find this even in extreme X-ray dominated regions \citep{spaans07,bayet11}. We therefore conclude that HCO$^+$ is self-absorbed in Zw049.057 and in IC860.

\section{Why does HCN-VIB thrive in obscured regions?}
\label{s:B}

The HCN $\nu_2$=1 vibrational ladder can commence being excited when $T_{\rm B}$(14$\mu$m)$\gapprox$100 K \citep{carroll81,aalto07}.  The mid-IR emission must therefore
be opaque when the dust temperature is $T_{\rm dust}$=100 K. This requires an optical depth of at least $\tau$(14$\mu$m)$\gapprox$5 and for a standard Galactic extinction curve
\citep{weingartner01} and $A_{\rm V} \approx N_{\rm H}/2\times 10^{21}$ $\cmmt$ this is equivalent to an H$_2$ column density $N$(H$_2$) $ \gapprox 2 \times 10^{23}$ $\cmmt$. 
For higher $T_{\rm dust}$ a lower $\tau$ will be sufficient for $T_{\rm B}$(IR)$=$100 K to be satisfied, but at a dust temperature of 400 K an H$_2$ column density in excess
of $=10^{23}$ $\cmmt$ is still necessary. 

The vibrational temperature $T_{\rm vib}$ is a measure of the HCN-VIB excitation temperature and hence of the brightness of the background mid-IR radiation field the molecules "see". 
We can estimate an effective $T_{\rm vib}$ through comparing the intensities of the HCN-VIB ($I_u$) and vibrational ground state lines ($I_l$) assuming optically thin emission, 

\begin{equation}
\label{e:vib}
I_u/I_l=A_u/A_l (\nu_u/\nu_l)^2 g_u/g_l \, {\rm exp}(-E_u/T_{\rm vib}). 
\end{equation}

\noindent
For a $T_{\rm vib}$ of 100 K the optically thin intensity ratio between the HCN-VIB and the HCN line is $3 \times 10^{-5}$ and the HCN-VIB line is not detectable. Thus, high column densities are
required to observe the HCN-VIB line when $T_{\rm vib}$=100 K.  In contrast, for $T_{\rm vib}$=400 K  the optically thin ratio between the HCN and the HCN-VIB line is
0.08. However, such a high $T_{\rm B}$(mid-IR) requires $T_{\rm dust}$=400 K combined with a large dust column density for optically thick emission - or a very large $T_{\rm dust}> 400 K$. 
Hence mid-IR excitation of HCN is favoured in regions of high dust temperature and with significant gas and dust column densities. 

\end{appendix}
\end{document}